\documentclass[options]{JHEP}

\title{Transgression forms and extensions of Chern-Simons gauge theories}
\author{Pablo Mora\\Instituto de F\'{\i}sica, Facultad de Ciencias, Igu\'a 4225,
Montevideo, Uruguay\\E-mail: pablmora-at-gmail-dot-com}
\author{Rodrigo Olea\\Pontificia Universidad
Cat\'olica de Chile, Casilla 306, Santiago 22, Chile
\\E-mail: rolea-at-chopin-dot-fis-dot-puc-dot-cl}
\author{Ricardo Troncoso\\Centro de Estudios Cient\'{\i}ficos CECS, Casilla 1469, Valdivia,
Chile
\\E-mail: ratron-at-cecs-dot-cl}
\author{Jorge Zanelli\\Centro de Estudios Cient\'{\i}ficos CECS, Casilla 1469, Valdivia, Chile
\\E-mail:jz-at-cecs-dot-cl}

\abstract{A gauge invariant action principle, based on the idea of
transgression forms, is proposed. The action extends the Chern-Simons form by
the addition of a boundary term that makes the action gauge invariant (and not
just quasi-invariant). Interpreting the spacetime manifold as cobordant to
another one, the duplication of gauge fields in spacetime is avoided. The
advantages of this approach are particularly noticeable for the gravitation
theory described by a Chern-Simons lagrangian for the AdS group, in which case
the action is regularized and finite for black hole geometries in diverse
situations. Black hole thermodynamics is correctly reproduced using either a
background field approach or a background-independent setting, even in cases
with asymptotically nontrivial topologies. It is shown that the energy found
from the thermodynamic analysis agrees with the surface integral obtained by
direct application of Noether's theorem.}

\begin{document}

\section{Introduction}

Unlike the theories for the other three known fundamental interactions, General
Relativity (GR), the currently accepted theory of gravitation described by the
Einstein-Hilbert action, is neither a gauge theory for the Poincar\'{e} nor the
(A)dS groups \footnote{For a discussion on this point, including references,
see e. g., \cite{QMFS,zanelli2}}. The reason is that while the spin connection
can be regarded as a gauge potential for the Lorentz group, the vielbein does
not transform as a connection for translations or (A)dS boosts.

Since the geometry is dynamically determined, the construction of a gauge
theory of gravity would require an action without reference to a fixed
space-time background. This requirement rules out actions of the Yang-Mills
type, which requires a background metric. It turns out that the only
possibility of a Lagrangian for gravity in terms of a connection without extra
fields, is given by the Chern-Simons (\textbf{CS}) form for some space-time
group (like the de Sitter, anti de Sitter or Poincar\'{e}
groups)\cite{QMFS,zanelli2}.

A CS gauge theory is one whose Lagrangian is given by the CS form for a gauge
group. These theories have been studied in a variety of contexts (see for
instance refs.\cite{aschwarz, deser,witten-cs}). In particular, CS gravities
and supergravities are defined by a space-time gauge group or one of its
supersymmetric extensions. These theories were introduced in refs.
\cite{vannieuwen,achu,witten1} for three-dimensional space-times. It was
noticed that General Relativity in 2+1 dimensions is a CS theory for the
Poincar\'{e} group, ISO(2,1), a fact that was exploited by Witten to show that
the theory is exactly solvable at the quantum level\cite{witten1}. There also
exist gravity theories in higher odd dimensions described in terms of CS
actions \cite{chamseddine, troncoso3}. For negative cosmological constant, the
local supersymmetric extension in five dimensions was given in
\cite{chamseddine-Nuc}, and for higher odd dimensions in \cite{troncoso1,
troncoso2}. In the absence of cosmological constant the corresponding
supergravity theories were constructed in \cite{banados1, HOT, hassaine}.

In \cite{troncoso2} it was also suggested that the low energy limit of M-theory
\cite{townsend1,hull1,witten2,townsend2} may be a CS theory with gauge group
$OSp(1|32)$, a proposal also explored in \cite{horava,banados2,nastase}.

For dimensions $d>3$, CS gravity theories are not equivalent to GR. The
question of the relationship between CS and GR in diverse dimensions has been
studied in refs.\cite{chamseddine,zanelli2, banados1, troncoso1, horava}.
Recently a new approach to this problem, and to the related one of finding a
\textquotedblleft non degenerate vacuum", has been discussed in ref.
\cite{hassaine} for the eleven-dimensional supergravity invariant under the M-algebra.

In this work we consider theories based on \textit{transgression forms}, which
are generalizations extending CS forms by the inclusion of a second gauge
field \cite{stora,zumino,manes,alvarez,chern,nakahara}. Conversely, CS forms
can be thought of as transgression forms with one of the gauge fields set
equal to zero. The second gauge field in the transgression form can be
interpreted either as a fixed, non-dynamical background, or as a dynamical
field on the same footing as the first one. In the second case, it is possible
to conceive both fields as defined on different manifolds with a common
boundary, thus eliminating possible ambiguities in the physical
interpretation. This is the point of view we advocate here.

Transgression forms can be used to define actions for physical systems that
give rise to well defined conserved charges \cite{potsdam,motz}, and in the
construction of actions for extended objects \cite{mn,mora}. More recently,
transgression forms have also been used in \cite{BFF2003,BFFF2005,IRS}.
Transgression forms in field theory were also the central topic of
ref.\cite{tesis}.

Using transgression forms to construct the action has several advantages:

(i) The transgression form singles out a unique boundary term for the action
principle, allowing both background-dependent and background-independent
formulations of the same system. In particular, in the background-independent
approach, as shown in Ref. \cite{motz}, it provides a well defined variational
principle for a wide set of boundary conditions.

(ii) In the case of gravitation, the boundary terms introduced by the
transgression allow to regularize the action for black hole configurations in
diverse situations. Thus, black hole thermodynamics is obtained using either a
background field approach or a background-independent setting, even in cases
with asymptotically nontrivial topologies. The results agree with the ones
computed by hamiltonian methods \cite{BTZentropy, scan, topo2}.

(iii) Conserved charges can be constructed as surface integrals through the
Noether method. The energy obtained in this way agrees with the result found
from thermodinamics.

The plan of this work is as follows. Section 2 is devoted to the construction
of lagrangians as transgression forms in which the two connection fields $A$
and $\overline{A}$ coexist in the same spacetime manifold, which is sufficient
to compute some quantities of physical interest. In Section 3 this
construction is applied to gravity with negative cosmological constant, in
order to make contact with the background-independent approach of \cite{motz}.
In section 4, the interpretation of Section 2 is revisited and a proposal is
advanced where the two fields $A$ and $\overline{A}$ are regarded as having
support in two distinct manifolds $M$ and $\overline{M}$ with a common
boundary. In Section 5, the new setting is applied to the spacetime geometry,
in particular for black holes of various dimensions and diferent topologies
verifying points (ii) and (iii) above. Section 6 contains the discussion and
comments, while reviewed material and some detailed calculations are contained
in the appendices.

\section{Transgression forms as Lagrangians}

A Chern-Simons form $\mathcal{C}_{2n+1}(A)$ is a differential form defined for
a connection $A$, whose exterior derivative yields a Chern class. Although the
Chern classes are gauge invariant, the CS forms are not; under gauge
transformations they change by a closed form. A transgression form
$\mathcal{T}_{2n+1}$ on the other hand, is an \emph{invariant} differential
form whose exterior derivative is the \emph{difference} of two Chern classes.
It generalizes the CS form with the additional advantage that it is gauge
invariant. The price to pay is that it is a function of two connections $A$ and
$\overline{A}$. In fact, a transgression form can be written as the difference
of two CS forms plus an exact form,
\begin{equation}
\mathcal{T}_{2n+1}=\mathcal{C}_{2n+1}(A)-\mathcal{C}_{2n+1}(\overline
{A})-dB_{2n}\left(  A,\overline{A}\right)  . \label{Transgression+CS+C2n}%
\end{equation}
It can be written as (see e.g., \cite{nakahara}),
\begin{equation}
\mathcal{T}_{2n+1}\left(  A,\overline{A}\right)  =(n+1)\int_{0}^{1}%
dt\ <\Delta{A}F_{t}^{n}>\ , \label{Transgression form}%
\end{equation}
where\footnote{Here wedge product between forms is assumed.}
\begin{eqnarray}
A_{t}  & =& tA+(1-t)\overline{A}\ \nonumber\label{At}\\
& =& \overline{A}+t\Delta A\ ,
\end{eqnarray}
is a connection that interpolates between the two independent gauge potentials
$A$ and $\overline{A}$. The Lie algebra-valued one-forms $A=A_{\mu}^{I}%
T_{I}\ dx^{\mu}$ and $\overline{A}=\overline{A}_{\mu}^{I}T_{I}\ dx^{\mu}$ are
connections under gauge transformations, $T_{I}$ are the generators, and
$<\cdots>$ stands for a symmetrized invariant trace in the Lie algebra (see
Appendix A). The corresponding curvature is
\begin{eqnarray}
F_{t}  & =& dA_{t}+A_{t}^{2}\ \nonumber\label{Ft}\\
& =& t F +(1-t)\overline{F}-t(1-t)(\Delta A)^{2}\ ,
\end{eqnarray}
and the explicit expression for the $2n$-form $C_{2n}$ is
\begin{equation}
B_{2n}=-n(n+1)\int_{0}^{1}ds\int_{0}^{1}dt~s~<A_{t}\Delta A~F_{st}^{n-1}>
\end{equation}
where $F_{st}=sF_{t}+s(s-1)A_{t}^{2}$. Hence, the role of the surface term
$B_{2n}$ is to cancel the variation of the bulk terms $\mathcal{C}_{2n+1}$,
which change by a closed form under a gauge transformation. The pure CS
density is recovered setting $\overline{A}=0$.

Transgression forms can be used as a field theory Lagrangians for gauge fields
$A$ and $\overline{A}$, where $B_{2n}$ is the interaction term which is only
defined at the boundary.

Conserved charges written as surface integrals for CS theories have been
obtained using different approaches in Refs. \cite{Silva,sarda}. Since the
transgression form (\ref{Transgression form}) is manifestly invariant under
diffeomorphisms and gauge transformations where both $A$ and $\overline{A}$
transform as connections simultaneously, the corresponding conserved charges
can be simply written as surface integrals by direct application of Noether's
theorem. Assuming suitable asymptotic conditions for the fields, one obtains
the conserved charges associated with an asymptotic Killing vector $\xi$ as
\begin{equation}
Q({\xi})=n(n+1)\int_{\partial\Sigma}\int_{0}^{1}dt<\Delta AF_{t}^{n-1}I_{\xi
}A_{t}>\ , \label{Qdiff}%
\end{equation}
where $\partial\Sigma$ is the boundary of the spatial section $\Sigma$.
Analogously, for an asymptotically covariantly constant Lie algebra valued
parameter $\lambda=\lambda^{I}T_{I}$, $D\lambda=0$ , the charges correspond
to
\begin{equation}
Q({\lambda})=n(n+1)\int_{\partial\Sigma}\int_{0}^{1}dt<\Delta AF_{t}%
^{n-1}\lambda>\ . \label{Qgauge}%
\end{equation}
The explicit derivation of (\ref{Qdiff}) and (\ref{Qgauge}) is simple
\cite{tesis,Thesis+Preprint+BFF+IRS}, and is reproduced here in Appendix B.

As explicit examples, the expression for the transgression form in three
dimensions is
\begin{equation}
\mathcal{T}_{3}=\mathcal{C}_{3}(A)-\mathcal{C}_{3}(\overline{A})-d<A\overline
{A}>\ ,
\end{equation}
where $\mathcal{C}_{3}(A)$ stands for the CS form
\begin{equation}
\mathcal{C}_{3}=<AF-\frac{1}{3}A^{3}>\;.
\end{equation}
Similarly, in five dimensions the transgression form turns out to be
\begin{equation}
\mathcal{T}_{5}(A,\overline{A})=\mathcal{C}_{5}(A)-\mathcal{C}_{5}%
(\overline{A})-dB_{4}(A,\overline{A})\ ,
\end{equation}
where the five-dimensional CS Lagrangian reads
\begin{equation}
\mathcal{C}_{5}=<AF^{2}-\frac{1}{2}A^{3}F+\frac{1}{10}A^{5}>\ ,
\end{equation}
and the boundary term is
\begin{equation}
B_{4}(A,\overline{A})=\frac{1}{2}<(A\overline{A}-\overline{A}A)(F+\overline
{F})+\overline{A}A^{3}+\overline{A}^{3}A+\frac{1}{2}A\overline{A}A\overline
{A}>\ .
\end{equation}

\section{Finite action for gravity for the AdS group}

Following the same basic principles of General Relativity in higher dimensions
allows for a wide class of gravity theories. The generalization of the
Einstein-Hilbert Lagrangian for any dimension $d$ are the so-called the
Lovelock Lagrangians \cite{Lovelock},\footnote{Latin indices $a,b$ run from
$0,...,d-1$.}
\begin{equation}
L_{Lovelock}=\kappa\int_{M}%
{\displaystyle\sum\limits_{p=0}^{n}}
\alpha_{p}L^{(p)}\ , \label{Lovelock}%
\end{equation}
where $L^{(p)}$ are the dimensional continuations of Euler densities from
lower dimensions
\[
L^{(p)}=\epsilon_{a_{1}\cdot\cdot\cdot a_{d}}R^{a_{1}a_{2}}\cdot\cdot\cdot
R^{a_{2p-1}a_{2p}}\ e^{a_{2p+1}}\cdot\cdot\cdot e^{a_{d}}\ .
\]
Here $e^{a}$ is the vielbein one-form, and $R^{ab}=d\omega^{ab}+\omega
_{\ c}^{a}\omega^{cb}$ is the curvature two-form.

\subsection{Chern-Simons gravity}

For $d=2n+1$ and the special choice of coefficients
\begin{equation}
\alpha_{p}=\frac{l^{2(p-n)}}{d-2p}\left(
\begin{array}
[c]{c}%
n\\
p
\end{array}
\right)  \ , \label{Coeffs}%
\end{equation}
the action corresponds to CS form for the AdS group \cite{chamseddine}, up to
a boundary term. It is useful to rewrite the series (\ref{Lovelock}) with the
choice (\ref{Coeffs}) as an integral over a continuous parameter $t\in
\lbrack0,1]$,
\begin{equation}
L_{CS}(\omega^{ab},e^{a})=\kappa\int_{0}^{1}dt\ \epsilon_{a_{1}\cdots
a_{2n+1}}R_{t}^{a_{1}a_{2}}\cdots R_{t}^{a_{2n-1}a_{2n}}\ e^{a_{2n+1}}\ ,
\label{Lcs}%
\end{equation}
with
\[
R_{t}^{ab}\equiv R^{ab}+t^{2}e^{a}e^{b}\ ,
\]
where the AdS radius $l$ has been set equal to one. Hereafter we choose
$\kappa=[2(d-2)!\Omega_{d-2}G]^{-1}$, where $\Omega_{d-2}$ is the volume of
the sphere $S^{d-2}$, and $G$ is the $d$-dimensional `Newton constant'.

The explicit link between this action and a CS form can be seen as follows.
The connection for the AdS group $SO(d-1,2)$ reads
\begin{equation}
A=\frac{1}{2}\omega^{ab}J_{ab}+e^{a}P_{a}\ ,
\end{equation}
where $J_{ab}$, and $P_{a}$ stand for the generators of Lorentz rotations and
AdS boosts, respectively. Here $e^{a}$ is identified with the vielbein and
$\omega^{ab}$ with the spin connection. The corresponding curvature
$F=dA+A^{2}$ is given by
\[
F=\frac{1}{2}\left(  R^{ab}+e^{a}e^{b}\right)  J_{ab}+T^{a}P_{a}\ ,
\]
where $T^{a}=De^{a}$ is the torsion two-form.

The AdS group admits an invariant tensor yielding the symmetric trace
\begin{equation}
<J_{a_{1}a_{2}}...J_{a_{2n-1}a_{2n}}P_{a_{2n+1}}>=\kappa\frac{2^{n}}%
{(n+1)}\epsilon_{a_{1}...a_{2n+1}}\ ,
\end{equation}
and it is simple to check that the Lagrangian in Eq.(\ref{Lcs})%
\[
L_{CS}(\omega^{ab},e^{a})=\kappa\int_{0}^{1}dt\ \epsilon\left(  R_{t}\right)
^{n}\ e\ ,
\]
satisfies\footnote{Here, for simplicity we have omitted the indices, which are
assumed to be contracted in canonical order.}
\begin{equation}
dL_{CS}=\kappa\epsilon(R+e^{2})^{n}T=<F^{n+1}>\ ,
\end{equation}
and hence, the Lagrangian is a CS form, up to an exact form.

\subsection{Transgression}

The explicit expression for the transgression form (\ref{Transgression form})
for the AdS group is obtained defining $\overline{A}=\frac{1}{2}%
\overline{\omega}^{ab}J_{ab}+\overline{e}^{a}P_{a}$ (see Appendix C), and is
written in terms of (\ref{Lcs}) as
\begin{equation}
\mathcal{T}_{2n+1}=L_{CS}(\omega,e)-L_{CS}(\overline{\omega},\overline
{e})-dB_{2n}\ . \label{Transgression-Gravity}%
\end{equation}
Since the transgression form is invariant by construction under local Lorentz
transformations, it cannot depend separately on $\omega$ and $\overline
{\omega}$, but only through the combination $\Delta\omega=\omega
-\overline{\omega}$, which transforms as a tensor. Indeed, the boundary term
is given by
\begin{equation}
B_{2n}=\kappa n\int_{0}^{1}dt\int_{0}^{1}ds~\epsilon~\Delta\omega
\ e_{t}\left[  tR+(1-t)\overline{R}-t(1-t)(\Delta\omega)^{2}+s^{2}e_{t}%
^{2}\right]  ^{n-1}\, \label{Transgression-gravity-BT}%
\end{equation}
where, from Eq. (\ref{At}), $e_{t}=te+(1-t)\overline{e}$.

Note that the Lorentz invariance of (\ref{Transgression-Gravity}) is manifest
since the curvatures, vielbeins and $\Delta\omega$ are Lorentz tensors.
However these are not tensors under AdS boosts and therefore, although the
full local AdS invariance is ensured by construction, this invariance is not
manifest in (\ref{Transgression-Gravity}).

An explicit example of (\ref{Transgression-Gravity}) and
(\ref{Transgression-gravity-BT}), in $d=2+1$ dimensions is the transgression
form
\begin{equation}
\mathcal{T}_{3}=\kappa\epsilon_{abc}(R^{ab}e^{c}+\frac{1}{3}e^{a}e^{b}%
e^{c})-\kappa\epsilon_{abc}(\overline{R}^{ab}\overline{e}^{c}+\frac{1}%
{3}\overline{e}^{a}\overline{e}^{b}\overline{e}^{c})-\kappa\frac{1}{2}%
\epsilon_{abc}d[\Delta\omega^{ab}(e^{c}+\overline{e}^{c})]\ ,
\end{equation}
which in a more compact notation reads
\begin{equation}
\mathcal{T}_{3}=\kappa\epsilon(Re+\frac{1}{3}e^{3})-\kappa\epsilon
(\overline{R}\overline{e}+\frac{1}{3}\overline{e}^{3})-\frac{1}{2}%
\kappa\epsilon d[\Delta\omega(e+\overline{e})]\ .
\end{equation}
Analogously, the transgression form in $d=4+1$ is
\begin{eqnarray}
\mathcal{T}_{5}  & =& \kappa\epsilon(R^{2}e+\frac{2}{3}Re^{3}+\frac{1}{5}
e^{5})-\kappa\epsilon(\overline{R}^{2}\overline{e}+\frac{2}{3}\overline
{R}\overline{e}^{3}+\frac{1}{5}\overline{e}^{5})\nonumber\\
& &-\frac{1}{3}\kappa\epsilon~d[\Delta\omega(e+\overline{e})(R-\frac{1}
{4}(\Delta\omega)^{2}+\frac{1}{2}e^{2})+\Delta\omega(e+\overline{e}
)(\overline{R}-\frac{1}{4}(\Delta\omega)^{2}+\frac{1}{2}\overline{e}^{2})\\
& &+\Delta\omega Re+\Delta\omega\overline{R}\overline{e}]\ .
\end{eqnarray}

In what follows, it is shown that the transgression form provides the suitable
boundary terms which yield well defined and finite action principles adapted
to different situations. Remarkably, regularized action principles using
background fields, as well as finite background-independent actions can be
obtained as particular cases within a unique framework. The thermodynamics of
black holes is then reproduced in both settings even in cases where the
horizon manifold has a nontrivial topology.

\subsection{Background-independent action and conserved charges}

A finite action principle that is background independent, must depend only on
the intrinsic geometric quantities, as the metric and the curvature, as well
as on the extrinsic curvature of the manifold at the boundary. This means that
apart from the vielbein and the curvature, the boundary term could depend only
on the second fundamental form, which is defined as
\[
\theta^{ab}=\omega^{ab}-\bar{\omega}^{ab}\ ,
\]
where $\bar{\omega}^{ab}$ is defined only at the boundary. Hence, $\bar
{\omega}^{ab}$ can be naturally identified with the spin connection of an
auxiliary manifold $\bar{M}$ which is cobordant with $M$, (i.e. $\partial
M=\partial\bar{M}$), and is endowed with a product metric which matches the
metric of $M$ at the boundary.

As a consequence, the required objects are the vielbein $e^{a}$, the spin
connection $\omega^{ab}$, as well as the auxiliary spin connection
$\bar{\omega}^{ab}$ chosen as described above. Such an action principle can
then be obtained through a transgression form (\ref{Transgression form}) with
\begin{eqnarray}
A  & =& \frac{1}{2}\omega^{ab}J_{ab}+e^{a}P_{a}\ ,\nonumber\\
\overline{A}  & =& \frac{1}{2}\bar{\omega}^{ab}J_{ab}\ \ .
\end{eqnarray}
It is worthwhile to point out that, since transgression form is invariant by
construction under local Lorentz transformations, it does not depend separately
on $\omega$ and $\bar{\omega}$, but only through the combination
$\Delta\omega^{ab}=\omega-\bar{\omega}=\theta^{ab}$, which transforms as a
tensor.

As announced in Ref. \cite{motz}, the required action principle is then
recovered by means of Eqs. (\ref{Transgression-Gravity}) and
(\ref{Transgression-gravity-BT}), so that the action becomes
\[
I=\int_{M}\mathcal{T}_{2n+1}\ ,
\]
with $\bar{e}^{a}=0$. In concrete, replacing $\bar{e}^{a}=0$ annihilates the
second bulk term in Eq. (\ref{Transgression-Gravity}). Making the same
replacement in (\ref{Transgression-gravity-BT}), using the Gauss-Codazzi
equation $R^{ab}=\bar{R}^{ab}+(\theta^{2})^{ab}$ for the relevant components
at the boundary, and changing $st\rightarrow s$, the boundary term turns out
to be
\[
B_{2n}=\kappa n\int_{0}^{1}dt\int_{0}^{t}ds~\epsilon~\theta\ e\left(  \bar
{R}+t^{2}\theta^{2}+s^{2}e^{2}\right)  ^{n-1}\ ,
\]
in agreement with \cite{motz}. It was shown that this boundary term is
sufficient to render the action finite for asymptotically locally AdS
solutions. Furthermore, the Euclidean continuation of the action correctly
describes the black hole thermodinamics in the canonical ensemble even in
cases with asymptotically nontrivial topology.

Following the same proceduce, the conserved charges written as surface
integrals obtained in \cite{motz}
\[
Q(\xi)=\ \kappa n\int_{\partial\Sigma}\int_{0}^{1}dt\ t\ \epsilon\left(
I_{\xi}\theta e+\theta I_{\xi}e\right)  \left(  \bar{R}+t^{2}\theta^{2}%
+t^{2}e^{2}\right)  ^{n-1},
\]
are recovered from Eq. (\ref{Qdiff}). It is worth mentioning that although the
black hole mass can be computed from two radically different approaches,
namely form thermodynamics or form the surface integrals, both results
completely agree, including the for the zero point (or Casimir) energy, which
corresponds to the mass of the locally AdS solutions.

\section{Reinterpretation of the theory}

The complete definition of the theory involves both the action principle and
suitable boundary conditions. For the action principle to be well defined the
action must have an extremum for solutions of the field equations satisfying
the boundary conditions.

The action could be taken to be just the integral on a single manifold $M$ of
the transgression form. This field theory would describe two self-interacting
fields $A$ and $\overline{A}$, which only interact with each other at the
boundary. This is a rather strange state of affairs: there is a duplicity of
identical dynamical fields which coexist in the spacetime $M$, but don't affect
each other, except by their interaction at the boundary However, since the
kinetic term for $\overline{A}$ has the wrong sign, this field would be a
\emph{phantom} with an ill-defined propagator.

One way to avoid this conflict is to assume $\overline{A}$ to be a non
dynamical background field. This action, however, would be gauge invariant
only up to a surface term.

There is a different conceptual framework where the transgression naturally
fits in.and which is free from the difficulties mentioned above, i. e., gauge
invariance and absence of phantoms. The idea is to conceive the fields $A$ and
$\overline{A}$ as defined on cobording manifolds $M$ and $\overline{M}$,
respectively, such that $\partial M\equiv\partial\overline{M}$. Then, we
propose the following action principle
\begin{equation}
I_{trans}=\int_{M}\mathcal{C}_{2n+1}(A)-\int_{\overline{M}}\mathcal{C}%
_{2n+1}(\overline{A})-\int_{\partial M}B_{2n}\left(  A,\overline{A}\right)
\,, \label{TransAct}%
\end{equation}
which describes two Chern-Simons systems interacting only at their common boundary.

Some remarks are in order:

\noindent(i) The action of Eq. (\ref{TransAct}) is exactly invariant under
gauge transformations since both $A$ and $\overline{A}$ transform as
connections in their respective domains, provided the gauge transformation is
continuous across $\partial M$.

\noindent(ii) The sign difference between the integrals on $M$ and on
$\overline{M}$ can be understood as due to the difference in orientations
necessary to match the two manifolds at their common boundary.

\noindent(iii) The action (\ref{TransAct}) is not the integral of the
transgression form on a single manifold with boundary.

What emerges is a field theory for two subsystems in contact at their common
boundary, each described by a CS lagrangian. This description is most natural
in the analysis of black hole thermodynamics below, where the two manifolds
$M$ and $\overline{M}$ don't even have the same topology (see, e.g., the
construction in \cite{motz}).

The variation of the action (\ref{TransAct}) is (see Appendix A)
\begin{equation}
\label{EOM}\delta I_{trans}=(n+1)\int_{M}<F^{n}\delta A>-\int_{\overline{M}%
}<\overline{F}^{n}\delta\overline{A}> +\int_{\partial M}\Theta_{2n}%
\end{equation}
The field equations,
\[
<F^{n}T_{I}>|_{M} =0\;,\;<\overline{F}^{n}T_{I}>|_{\overline{M}} =0\ \ ,
\]
coincide with those of a pure CS theory for two independent connections in the
corresponding manifolds $M$ and $\overline{M}$. The action attains an extremum
provided the boundary term
\begin{equation}
\Theta_{2n}=-n(n+1)\int_{0}^{1}<\Delta AF_{t}^{n-1}\delta A_{t}>
\label{Theta2n}%
\end{equation}
vanishes on $\partial M$. A sufficient boundary condition would be, for
instance, to require $\Delta A\rightarrow0$ at the boundary, with a fast
enough fall-off in the direction normal to the boundary, while $F_{t}$ remains
finite at the boundary so that $\Theta_{2n}=0$. One may call this case where
the connection $A$ approaches a reference field configuration $\overline{A}$
at the boundary, \textit{background dependent}. Alternatively, the approach in
which $A$ and $\overline{A}$ are both dynamical fields, can be called
\textit{background independent}. Of course, there exist infinitely many other
ways to ensure the vanishing of (\ref{Theta2n}), in which some components
approach a reference connection, while others don't, for example. We shall
make use of this possibility in Sect. 3.1 below.

In varying the action, one could also assume that $A$ is dynamical, while
$\overline{A}$ is a fixed background which should not be varied. In this case,
the second term in the R.H.S. of (\ref{EOM}) wouldn't exist and only $A$ needs
to satisfy the field equations. In any case, $\overline{A}$ could also be
taken as a special solution of the field equations, identified as a
\textquotedblleft vacuum". However this means that the canonical realization
of gauge invariance may break down at the boundary.

\section{Application to gravity}

\subsection{Black hole thermodynamics with a reference geometry}

The purpose of this subsection is to show that the same transgression form
(\ref{Transgression form}) also provides an alternative way to obtain a
regularized action with a reference background geometry ($\overline{e}^{a}%
\neq0$), which may work in more exotic situations.

The action principle we now consider is
\begin{equation}
I_{trans}=\int_{M}L_{CS}(\omega,e)-\int_{\overline{M}}L_{CS}(\overline{\omega
},\overline{e})-\int_{\partial M}B_{2n}\ , \label{Ireg}%
\end{equation}
where $L_{CS}$ is the Lagrangian defined by Eq. (\ref{Lcs}), and $B_{2n}$ is
the boundary term (\ref{Transgression-gravity-BT}). Here the spin connection
$\omega^{ab}$ and the vielbein $e^{a}$ have support only on the manifold $M$,
while $\overline{\omega}^{ab}$ and $\overline{e}^{a}$ have support only on the
cobordant manifold $\overline{M}$.

In what follows it is shown that in the Euclidean continuation, the
regularized action (\ref{Ireg}) is finite and gives the correct free energy in
the canonical ensemble for black holes, even in cases with nontrivial
topology. It is also shown that the energy found from the thermodynamic
analysis coincides with the one obtained from the Noether theorem.

\subsubsection{Euclidean action and black hole thermodynamics}

We consider a family of black holes whose horizons may have a non-spherical
topology, labeled by the parameter $\gamma$ which can take the values $\pm
1,0$. The line element is \cite{dimensionally,topo1,topo2,scan}
\begin{equation}
ds^{2}=-\Delta^{2}dt^{2}+\frac{1}{\Delta^{2}}dr^{2}+r^{2}d\Sigma_{d-2}^{2}\ ,
\label{BH-Solution}%
\end{equation}
with%
\begin{equation}
\Delta^{2}=\gamma-\sigma+r^{2}\ ,
\end{equation}
where $d\Sigma_{d-2}^{2}$ is the line element of the $(d-2)$-dimensional base
manifold of constant curvature proportional to $\gamma=1,0,-1$. The horizon is
located at $r_{+}=\sqrt{\sigma-\gamma}$, and in the euclidean continuation, the
manifold for a massive black hole has a radial coordinate that extends over the
range $r_{+}\leq r<\infty$. The euclidean time period $\beta$ which determines
the temperature is found demanding smoothness of the Euclidean
solution at the horizon%
\[
\beta=T^{-1}=\frac{2\pi}{r_{+}}\ .
\]
For a fixed temperature, in the semiclassical approximation, the Euclidean
action is related to the free energy $F$ in the canonical ensemble,
$I_{E}=-\beta F=-\beta E+S$.

Here it is shown that the black hole thermodynamics is reproduced evaluating
the solution in euclidean continuation of the action in Eq. (\ref{Ireg}).
Thus, we consider $A$ corresponding to a black hole solution of the form
(\ref{BH-Solution}), while $\overline{A}$ is assumed to be a reference
configuration given by a suitable solution of the field equations. Since the
time period $\beta$ is fixed for the Euclidean black hole solution, the
reference background must be such that $\bar{\beta}$ is arbitrary in order to
have a well-defined cobordism between the manifolds $M$ and $\bar{M}$. This
requirement is fulfilled for the solution (\ref{BH-Solution})) with $r_{+}=0$,
i.e., for $\bar{\sigma}=\gamma$, as well as for AdS spacetime which
corresponds to $\bar{\sigma}=0$ in the spherically symmetric case ($\gamma=1$).

\textbf{A. Asymptotic spherical symmetry ($\gamma=1$)}

In this case there are two possible reference geometries with arbitrary
$\beta$, AdS space and the \textquotedblleft black hole vacuum" with
$\bar{\sigma}=1$. In both cases the range of the radial coordinate is $0\leq
r<\infty$.

The bulk contributions to the euclidean continuation of the action
(\ref{Ireg}) are
\begin{eqnarray}
I_{trans}^{bulk}  & =& \kappa\beta(d-2)!\Omega_{d-2}\int_{0}^{1}ds\left.
[(\sigma+(s^{2}-1)r^{2})^{n}+2nr^{2}(s^{2}-1)(\sigma+(s^{2}-1)r^{2}
)^{n-1}]\right\vert _{r=r_{+}}^{r=\infty}\nonumber\\
& & -\kappa\beta(d-2)!\Omega_{d-2}\int_{0}^{1}ds\left.  [(\overline{\sigma
}+(s^{2}-1)r^{2})^{n}+2nr^{2}(s^{2}-1)(\overline{\sigma}+(s^{2}-1)r^{2}%
)^{n-1}]\right\vert _{r=0}^{r=\infty},\nonumber
\end{eqnarray}
where $\Omega_{d-2}$ is the volume of the ${d-2}$-dimensional unit sphere. The
boundary term is
\begin{eqnarray}
\int_{\partial\mathcal{M}}B_{2n} =& &-2n\kappa\beta(d-2)!\Omega_{d-2}%
\lim_{r\rightarrow\infty}\int_{0}^{1}dt\int_{0}^{1}ds\{(\Delta-\overline
{\Delta})[t\Delta+(1-t)\overline{\Delta}]\times\nonumber\\
& & \times\lbrack1-(t\Delta+(1-t)\overline{\Delta})^{2}+s^{2}r^{2}%
]^{n-1}\nonumber\\
& & +2(n-1)r^{2}(s^{2}-1)(\Delta-\overline{\Delta})[t\Delta+(1-t)\overline
{\Delta}][1-(t\Delta+(1-t)\overline{\Delta})^{2}+s^{2}r^{2}]^{n-2}\}\nonumber
\end{eqnarray}
Integration on $t$ can be performed with the substitution $u=1-(t\Delta
+(1-t)\overline{\Delta})^{2}+s^{2}r^{2}$.

Hence, the boundary term exactly cancels the divergent contributions of the
bulk (corresponding to the $r\rightarrow\infty$ limit), so that the total
action (\ref{Ireg}) is finite and given by
\begin{equation}
I_{trans}=\frac{\beta n}{G}r_{+}\int_{0}^{r_{+}}dx(1+x^{2})^{n-1}-\frac{\beta
}{2G}(\sigma^{n}-\bar{\sigma}^{n})\;. \label{Action-spherically-symmetric}%
\end{equation}
The entropy is then given by%
\begin{equation}
S =[1-\beta\frac{\partial~}{\partial\beta}]I_{trans}\\
=\frac{2\pi n}{G}\int_{0}^{r_{+}}dx(1+x^{2})^{n-1}\ ,
\end{equation}
which doest not depend on the choice of reference configuration in agreement
with previous calculations done by other methods
\cite{dimensionally,scan,motz}. The energy is
\begin{equation}
E =-\frac{\partial I_{trans}}{\partial\beta}\ =\frac{1}{2G}(\sigma^{n}%
-\bar{\sigma}^{n})\ ,
\end{equation}
which depends on the choice of reference background. Note that if the
reference configuration is taken to be the black hole vacuum ($\bar{\sigma}%
=1$) these results agree with the ones of Ref. \cite{motz}, where AdS
spacetime can be regarded to have a nonvanishing \textquotedblleft Casimir"
energy given by $E_{AdS}=-(2G)^{-1}$.

\textbf{B. Other topologies ($\gamma=0,-1$)}

The evaluation of the euclidean action for black holes with nontrivial
topology follows the same steps as in the previous case. The background
configuration is the one with $r_{+}=0$ which corresponds to choose
$\bar{\sigma}=\gamma$. Again, the divergent part of the bulk contributions
cancel the boundary term, and now the action becomes
\begin{equation}
I_{trans}=\frac{\beta n}{G}\frac{\Sigma_{d-2}}{\Omega_{d-2}}r_{+}\int
_{0}^{r_{+}}dx(\gamma+xr^{2})^{n-1}-\frac{\beta}{2G}\frac{\Sigma_{d-2}}%
{\Omega_{d-2}}(\sigma^{n}-\gamma^{n})\;,
\end{equation}
where $\Sigma_{d-2}$ stands for the volume of the $d-2$ dimensional base
manifold. The entropy now reads
\begin{equation}
S=\frac{2\pi n}{G}\frac{\Sigma_{d-2}}{\Omega_{d-2}}\int_{0}^{r_{+}}%
dx(\gamma+x^{2})^{n-1}\ ,
\end{equation}
and the energy is \ given by%
\begin{equation}
E=\frac{\Sigma_{d-2}}{\Omega_{d-2}}(\sigma^{n}-\gamma^{n})\ .
\end{equation}
Note as in the background independent approach, the energy of the
configuration with negative constant curvature ($\sigma=0$) depends on the
topology and is given by $E=\gamma^{n}(2G)^{-1}$ in agreement with \cite{motz}.

Note that the background substraction procedure that occurs here is not the
same as the one proposed by Gibbons and Hawking \cite{gh}, or by Hawking and
Page \cite{hp}. In those papers, the actions for two different configurations
(for instance, for a massive black hole and Minkowski space) are subtracted,
with the additional condition that the metrics match at a very large finite
radius $r_{0}$ (eventually taken to infinity). That implies two different
Euclidean time intervals $\beta$ and $\overline{\beta}$. Although
$\overline{\beta}\rightarrow\beta$ when $r_{0}\rightarrow\infty$, there is an
extra contribution to the bulk action coming from the difference of the
$\beta$'s. In our approach there is always only one $\beta$, as it must be in
order to integrate the boundary term $B_{2n}$, where both sets of vielbein and
spin connections appear entangled, and we have an extra contribution coming
from that boundary term. The boundary terms obtained in the standard way
coincide with our approach for 2+1 dimensions only.

\subsection{Noether charges for AdS gravity}

In this section, we show that the energy found from the thermodynamic analysis
discussed above agrees with the computation from direct application of
Noether's theorem.

\subsubsection{Black hole mass from the asymptotic timelike isometry}

The Noether charge associated to isometries generated by a vector $\xi^{\mu}$
is
\begin{equation}
Q(\xi)=n(n+1)\int_{\partial\Sigma}\int_{0}^{1}dt<\Delta AF_{t}^{n-1}I_{\xi
}A_{t}>\ . \label{Qchi}%
\end{equation}
Here $A$ and $\overline{A}$ correspond to two arbitrary configurations of the
form (\ref{BH-Solution}) for the same topology at the boundary, i. e., for the
same $\gamma$. For the timelike Killing vector $\xi=\partial~/\partial t$,
(\ref{Qchi}) gives (see Appendix D for the details)%
\begin{equation}
Q\left(  \frac{\partial~}{\partial t}\right)  =\frac{\Sigma_{d-2}}%
{\Omega_{d-2}}(\sigma^{n}-\bar{\sigma}^{n})=E-\bar{E}\ ,
\end{equation}
where $E$ stands for the energy computed from thermodynamics. Thus, if
$\overline{A}$ is chosen as a reference background solution of the previous
section, this charge reproduces the energy computed above. This expression
also coincides with the result obtained from the background independent
approach in Ref. \cite{motz} when the background configuration is the one for
which the horizon radius vanishes, i.e., choosing $\bar{\sigma}=\gamma$.

\subsubsection{Black hole mass from the gauge Noether charge}

Alternatively, the energy can be obtained from the Noether charge associated
to gauge transformations, which is given by
\begin{equation}
Q({\lambda})=n(n+1)\int_{\partial\Sigma}\int_{0}^{1}dt<\Delta AF_{t}%
^{n-1}\lambda>\ . \label{QLBH}%
\end{equation}
The charge is then evaluated taking both $A$ and $\overline{A}$ as two
arbitrary solutions of the the form (\ref{BH-Solution}), with the same
topology at the boundary, for an asymptotically covariantly constant gauge
parameter $\lambda=\lambda^{a}P_{a}+\frac{1}{2}\lambda^{ab}J_{ab}$ satisfying
$\delta_{\lambda}A=D\lambda=d\lambda+A\lambda-\lambda A=0$ for $r\rightarrow
\infty$.

The identity $\mathcal{L}_{\xi}A=D(I_{\xi}A)+I_{\xi}F$, allows to identify the
Lie algebra valued parameter $\lambda$ with a Killing vector as $\lambda
=I_{\xi}A$ provided the curvature vanishes sufficiently fast at infinity. Thus,
choosing the gauge parameter as

\begin{equation}
\lambda=I_{\xi}A_{r\rightarrow\infty} \rightarrow rP_{0} +rJ_{01}\ .
\label{Lambda}%
\end{equation}
where $\xi$ is the timelike Killing vector $\partial~/\partial t$ allows to
obtain the the difference between the energies from (\ref{QLBH})
\[
Q\left(  \lambda\right)  =\frac{\Sigma_{d-2}}{\Omega_{d-2}}(\sigma^{n}%
-\bar{\sigma}^{n})=E-\bar{E}\ .
\]
In the spherically symmetric case, black hole solutions possess $d-1$
independent solutions depending on the arbitrary constants $C^{1}$, $C^{m}$,
with $m=2,...,d-1$, given by%
\begin{eqnarray}
\lambda^{1}  & =& \lambda^{0m}=0\\
\lambda^{0}  & =& \lambda^{01}=C^{1}r\\
\lambda^{m}  & =& \lambda^{m1}=C^{m}r\\
\lambda_{n}^{m}\tilde{e}^{n}  & =& \omega_{n}^{m}C^{n}\ ,
\end{eqnarray}
so that (\ref{Lambda}) is recovered choosing the parameters as $C^{1}=1$ and
$C^{m}=0$.

\section{Discussion and Comments}

The results reported here (and in ref.\cite{motz}) support the conjecture that
the boundary terms dictated by gauge invariance, supplemented by boundary
conditions that make the action principle well defined, give the right
conserved charges and black hole entropy without requiring any regularization.
The similar problem of computing the conserved charges for the Lovelock
theories of gravity in even dimensions was studied in ref.\cite{aroscargas},
where it was shown that it is possible to regularize the action and the
charges by adding a surface term whose exterior derivative is the Euler
density of the spacetime manifold.

As we mentioned, $\overline{A}$ could be regarded as a fixed reference or
background configuration, a non dynamical entity. However, a better option is
to assume both $A$ and $\overline{A}$ in different manifolds with a common
boundary. The calculation of the entropy with $\overline{A}$ corresponding to
AdS or to a zero mass black hole supports this option. About the question of
what is dynamical and what isn't it is worthwhile to remember the comment by
Yang and Mills in their landmark paper on non abelian gauge fields about the
need to give dynamical content to the gauge field, which they called $B_{\mu}$
\cite{yang-mills}.

It would be interesting to explore the application of the ideas presented for
CS gravity in the presence of \textquotedblleft exotic" parity-violating terms
\cite{witten1,Mardones-z,troncoso3}, as well as for its supersymmetric
extensions, and in the framework of the AdS/CFT\ correspondence
\cite{maldacena2}.

The lesson from this discussion is that the action (\ref{TransAct}), inspired
by the transgression form, has an important advantage over the pure CS action.
The boundary term incorporated in this way renders the action principle well
defined and the Euclidean action for black holes finite, whereas the pure CS
action diverges. Hence, it is natural to think of the boundary terms as
regulator for the CS theory. It is somehow surprising that these difficulties
are solved in one stroke just by requiring strict gauge invariance, and this
also suggests that the action principle defined by (\ref{TransAct}), would be
a better starting point for a path integral quantization.

\acknowledgments{We thank S. Willison for useful discussions. P.M. is grateful
for the hospitality at CECS-Valdivia, for multiple visits while this work was
being done. R.T. thanks the Universidad de la Rep\'{u}blica, Uruguay, for
hospitality during part of the writing of this paper. This work is partially
funded by FONDECYT grants 1020629, 1040921, 1051056, 1061291, 3030029, and
3040026. The generous support to CECS by Empresas CMPC is also acknowledged.
CECS is a Millenium Science Institute and is funded in part by grants from
Fundaci\'{o}n Andes and the Tinker Foundation.} \newline

\centerline{\textbf{APPENDICES}}

\appendix

\section{Invariant polynomials, transgression forms and CS forms}

In this appendix we review for completeness some elements of the theory of
fiber bundles used in this paper. This material can be found in
refs.\cite{stora,zumino,manes,alvarez,nakahara,bertlmann}. We are not aware of
any reference for the explicit formula for the variation of the transgression
that we used (though it is probably known) so we give a derivation in the last
subsection of this appendix.\newline An \textit{invariant polynomial} $~P(F)$~
is defined as the formal sum
\begin{equation}
P(F)=\sum_{n=0}^{N}\alpha_{n}<F^{n+1}>
\end{equation}
where
\[
<T_{I_{1}}\dots T_{I_{n+1}}>=g_{I_{1}\cdots I_{n+1}}
\]
corresponds to a \textit{symmetric invariant trace} in the algebra of $G$. This
is equivalent to say that $g_{I_{1}\cdots I_{n+1}}$ is an invariant symmetric
tensor in the algebra of $G$, which by construction has its indices in the
adjoint representation of $G$.

It can be shown that the invariant polynomials are closed
\[
dP(F)=0
\]
therefore locally exact
\[
P(F)=d\mathcal{C}_{2n+1}(A,F)
\]
where we introduced the CS form defined by
\[
\mathcal{C}_{2n+1}(A,F)\equiv(n+1)\int_{0}^{1}ds~\,<AF_{s}^{n}>
\]
with $A_{s}=sA$ and $F_{s}=dA_{s}+A_{s}^{2}$.

A similar relation which holds globally is the \textit{transgression formula},
involving two potentials $A$ and $\overline{A}$ in the same fiber, with
curvatures $F$ and $\overline{F}$ respectively
\[
<F^{n+1}>-<\overline{F}^{n+1}>=d\mathcal{T}_{2n+1}(A,\overline{A})
\]
with the \textit{transgression form} defined as
\[
\mathcal{T}_{2n+1}(A,\overline{A})\equiv(n+1)\int_{0}^{1}dt~\,<(A-\overline
{A})F_{t}^{n}>
\]
with $A_{t}=tA+(1-t)\overline{A}$ and $F_{t}=dA_{t}+A_{t}^{2}$.

The transgression form is invariant under gauge transformations for which $A$
and $\overline{A}$ transform with the same group element $g$ of the group $G$,
due to the covariance of $\Delta A\equiv A-\overline{A}$, $(\Delta
A)^{g}=g^{-1}(\Delta A)g$, the covariance of $F_{t}$, $F_{t}^{g}=g^{-1}F_{t}g
$, and the invariance of the symmetrized trace.

\subsection{Cartan operator and homotopy formula}

Let $A_{t}$ be the interpolation between two gauge potentials $A$ and
$\overline{A}$,
\[
A_{t}=tA+(1-t)\overline{A}~~,~~F_{t}=dA_{t}+A_{t}^{2}~~.
\]
The \textit{Cartan homotopy operator} $k_{01}$ acts on polynomials
$\mathcal{P}(F_{t},A_{t})$ and is defined as
\[
k_{01}\mathcal{P}(F_{t},A_{t})=\int_{0}^{1}dt~l_{t}\mathcal{P}(F_{t}%
,A_{t})~~,
\]
where the action of the operator $l_{t}$ on arbitrary polynomials of $A_{t}$
and $F_{t}$ is defined through
\[
l_{t}A_{t}=0~~,~~~~l_{t}F_{t}=A-\overline{A}\equiv\Delta A~~,
\]
and the convention that $l_{t}$ acts as an antiderivative $l_{t}(\Lambda
_{p}\Sigma_{q})=(l_{t}\Lambda_{p})\Sigma_{q}+(-1)^{p}\Lambda_{p}(l_{t}%
\Sigma_{q})$, where $\Lambda_{p}$ and $\Sigma_{q}$ are $p$ and $q$-forms
(polynomials in $A_{t}$ and $F_{t}$) respectively.

It can be verified the relationship
\[
\big(l_{t}d+dl_{t}\big)\mathcal{P}(F_{t},A_{t})=\frac{\partial}{\partial
t}\mathcal{P}(F_{t},A_{t})
\]
which can be integrated between 0 and 1 in $t$ to obtain the \textit{Cartan
homotopy formula}
\[
\big(k_{01}d+dk_{01}\big)\mathcal{P}(F_{t},A_{t})=\mathcal{P}(F,A)-\mathcal{P}%
(\overline{F},\overline{A})~~.
\]
For $\mathcal{P}=<F^{n+1}>$ we recover the transgression formula. Putting
$\mathcal{P}=\mathcal{C}_{2n+1}$ we get
\[
\mathcal{T}_{2n+1}=\mathcal{C}_{2n+1}(A,F)-\mathcal{C}_{2n+1}(\overline
{A},\overline{F})-d[k_{01}\mathcal{C}_{2n+1}]
\]
The $2n$ form $B_{2n}$ is defined by
\[
B_{2n}(A,F;\overline{A},\overline{F})\equiv k_{01}\mathcal{C}_{2n+1}%
\]
Explicitly
\begin{equation}
B_{2n}=-n(n+1)\int_{0}^{1}ds\int_{0}^{1}dt~s~<A_{t}\Delta A~F_{st}^{n-1}>
\end{equation}
where $F_{st}=sF_{t}+s(s-1)A_{t}^{2}$

\subsection{General variation of the transgression}

The transgression form is
\[
\mathcal{T}_{2n+1}=(n+1)\int_{0}^{1}dt<\Delta AF_{t}^{n}>
\]
with $\Delta A=A-\overline{A}$. furthermore
\[
A_{t}=t\Delta A+\overline{A}=tA+(1-t)\overline{A}%
\]
and
\[
F_{t}=dA_{t}+A_{t}^{2}=\overline{F}+t\overline{D}(\Delta A)+t^{2}(\Delta
A)^{2}%
\]
with $\overline{F}=d\overline{A}+\overline{A}^{2}$ and $\overline{D}(\Delta
A)=d(\Delta A)+\overline{A}(\Delta A)+(\Delta A)\overline{A}$. Notice that the
derivative of $F_{t}$ with respect to the parameter $t$ satisfy
\[
\frac{d~}{dt}F_{t}=D_{t}(\Delta A)=d(\Delta A)+A_{t}(\Delta A)+(\Delta
A)A_{t}=d(\Delta A)+2t(\Delta A)^{2}+\overline{A}(\Delta A)+(\Delta
A)\overline{A}%
\]
For the general variation of the transgression form we have
\[
\delta\mathcal{T}_{2n+1}=(n+1)\int_{0}^{1}dt\{<F_{t}^{n}\delta(\Delta
A)>+<n(\Delta A)F_{t}^{n-1}D_{t}[\delta A_{t}]>\}
\]
but, inside the bracket,
\[
D_{t}[\Delta AF_{t}^{n-1}\delta A_{t}]=D_{t}(\Delta A)F_{t}^{n-1}\delta
A_{t}-\Delta AF_{t}^{n-1}D_{t}[\delta A_{t}]=\frac{d~}{dt}F_{t}F_{t}%
^{n-1}\delta A_{t}-\Delta AF_{t}^{n-1}D_{t}[\delta A_{t}]
\]
and using $\delta A_{t}=t\delta(\Delta A)+\delta\overline{A}$ we get
\[
\delta\mathcal{T}_{2n+1}=(n+1)\int_{0}^{1}dt\{<[F_{t}^{n}+tn\frac{d~}{dt}%
F_{t}F_{t}^{n-1}]\delta(\Delta A)>+<n\frac{d~}{dt}F_{t}F_{t}^{n-1}%
\delta\overline{A}>\}
\]
\[
-n(n+1)~d~\int_{0}^{1}dt<\Delta AF_{t}^{n-1}\delta A_{t}>
\]
but , when inside the bracket, $F_{t}^{n}+tn\frac{d~}{dt}F_{t}F_{t}%
^{n-1}=\frac{d~}{dt}[tF_{t}^{n}]$ and $n\frac{d~}{dt}F_{t}F_{t}^{n-1}%
=\frac{d~}{dt}F_{t}^{n}$, which allow to evaluate explicitly the first to
integrals in $t$ giving
\[
\delta\mathcal{T}_{2n+1}=(n+1)<F^{n}\delta(\Delta A)>+(n+1)<(F^{n}%
-\overline{F}^{n})\delta\overline{A}>-n(n+1)~d~\int_{0}^{1}dt<\Delta
AF_{t}^{n-1}\delta A_{t}>
\]
and finally we get for generic infinitesimal variations of the transgressions
\[
\delta\mathcal{T}_{2n+1}=(n+1)<F^{n}\delta A>-(n+1)<\overline{F}^{n}%
\delta\overline{A}>-n(n+1)~d~\int_{0}^{1}dt<\Delta AF_{t}^{n-1}\delta A_{t}>
\]

\section{Noether's theorem and Conserved Charges}

\subsection{Noether's Theorem}

The variation of differential forms under diffeomorphisms for which the
coordinates change as $\delta x^{\mu}=\xi^{\mu}$ is given by
\[
\delta\alpha(x)=\alpha^{\prime}(x)-\alpha(x)=-\mathcal{L}_{\xi}\alpha
\]
where $\mathcal{L}_{\xi}$ is the Lie derivative, which for differential forms
can be written as
\[
\mathcal{L}_{\xi}\alpha=[dI_{\xi}+I_{\xi}d]\alpha
\]
with $d$ the exterior derivative and the contraction operator given by
\[
I_{\xi}\alpha_{p}=\frac{1}{(p-1)!}\xi^{\nu}\alpha_{\nu\mu_{1}...\mu_{p-1}%
}dx^{\mu_{1}}...dx^{\mu_{p-1}}%
\]
The operator $I_{\xi}$ is an antiderivation, in the sense that acting on the
exterior product of two differential forms $\alpha_{p}$ and $\beta_{q}$ of
orders $p$ and $q$ it gives $I_{\xi}(\alpha_{p}\beta_{q})=I_{\xi}\alpha
_{p}\beta_{q}+(-1)^{p}\alpha_{p}I_{\xi}\beta_{q}$. A useful result is that the
Lie derivative acting on gauge potentials is
\[
\mathcal{L}_{\xi}A=D(I_{\xi}A)+I_{\xi}F
\]
where $D$ is the covariant derivative and $F$ the field tensor.\newline One
consider a lagrangian density given by a differential form $L(\phi
,\partial\phi)$, where $\phi$ represents all the dynamical fields. The
variation of the lagrangian under diffeomorphisms is given by $\delta
L=-d(I_{\xi}L)$, as $dL=0$ because the order of $L$ is equal to the dimension
of the space. One considers a class of transformations under which the
lagrangian is quasi-invariant, combined with diffeomorphisms. Under these
transformations the variation of the lagrangian is
\[
\delta L=d\Omega-d(I_{\xi}L)
\]
where the first total derivative come from the transformations considered and
the second one from the diffeomorphisms.

On the other hand, the standard procedure leading to the equations of motion
gives the variation of the lagrangian as the equations of motion times the
variation of $\phi$ plus a boundary term
\[
\delta L=(E.d.M.)\delta\phi+d\Theta
\]
where the variations $\delta\phi$ are infinitesimal but arbitrary in form.
>From this two expressions of the variation, assuming that the variations in
both are restricted to transformations of the class considered in the first
expression of $\delta L$ and equating, that if the E.O.M. hold
\[
d[\Omega-I_{\xi}L-\Theta]=0
\]
It follows that the so called Noether current
\[
\star j=\Omega-I_{\xi}L-\Theta
\]
is conserved $d(\star j)=0$.\newline

In the next two subsections we will deduce the general form of the gauge and
diffeomorphism Noether charges for Transgression and Chern-Simons theories. In
the calculation of the expressions for the charges both gauge fields appearing
in the transgression are varied. If one prefers to consider the second field
as non dynamical, one could think of varying it as a trick, analogous to
varying the flat Minkowski metric to compute the energy momentum tensor of a
field in flat space-time, for a theory for which the metric is non dynamical.

\subsection{Diffeomorphism Noether charges}

The variation of the gauge potentials under diffeomorphisms is
\begin{eqnarray}
\delta_{\xi}A  & =& -\mathcal{L}_{\xi}A=-D[I_{\xi}A]-I_{\xi}F=-[I_{\xi
}d+dI_{\xi}]A\\
\delta_{\xi}\overline{A}  & =& -\mathcal{L}_{\xi}\overline{A}=-\overline
{D}[I_{\xi}\bar{A}]-I_{\xi}\bar{F}=-[I_{\xi}d+dI_{\xi}]\overline{A}\\
\delta_{\xi}A_{t}  &=& -\mathcal{L}_{\xi}A_{t}=-D_{t}[I_{\xi}A_{t}]-I_{\xi
}F_{t}=-[I_{\xi}d+dI_{\xi}]A_{t}
\end{eqnarray}
Inserting this in the variation of the transgression
\[
\delta\mathcal{T}_{2n+1}=(n+1)<F^{n}\delta A>-(n+1)<\bar{F}^{n}\delta\bar
{A}>-n(n+1)~d~\int_{0}^{1}dt<\Delta AF_{t}^{n-1}\delta A_{t}>
\]
we can read the form $\Theta$ that appears in the Noether theorem
\begin{equation}
\Theta=-n(n+1)\int_{0}^{1}dt<\Delta AF_{t}^{n-1}\delta_{\xi}A_{t}>
\end{equation}
or
\begin{equation}
\Theta=n(n+1)\int_{0}^{1}dt<\Delta AF_{t}^{n-1}D_{t}[I_{\xi}A_{t}]+\Delta
AF_{t}^{n-1}I_{\xi}F_{t}>
\end{equation}
But, inside the bracket,
\[
D_{t}[\Delta AF_{t}^{n-1}I_{\xi}A_{t}]=D_{t}\Delta AF_{t}^{n-1}I_{\xi}%
A_{t}-\Delta AF_{t}^{n-1}D_{t}[I_{\xi}A_{t}]=\frac{d~}{dt}F_{t}F_{t}%
^{n-1}I_{\xi}A_{t}-\Delta AF_{t}^{n-1}D_{t}[I_{\xi}A_{t}]
\]
then
\begin{equation}
\Theta=n(n+1)\int_{0}^{1}dt<\frac{d~}{dt}F_{t}F_{t}^{n-1}I_{\xi}A_{t}+\Delta
AF_{t}^{n-1}I_{\xi}F_{t}>-n(n+1)~d~\int_{0}^{1}dt<\Delta AF_{t}^{n-1}I_{\xi
}A_{t}>
\end{equation}
For the term $I_{\xi}L$ in the Noether current we have
\begin{equation}
I_{\xi}L=I_{\xi}\mathcal{T}_{2n+1}=(n+1)\int_{0}^{1}dt<I_{\xi}(\Delta
A)F_{t}^{n}-n\Delta AF_{t}^{n-1}I_{\xi}F_{t}>
\end{equation}
The current is $\ast j=\Omega-[\Theta+I_{\xi}L]$, but $\Omega=0$ because the
action is invariant under diffeomorphisms, then
\[
\ast j=-[\Theta+I_{\xi}L]=-(n+1)\int_{0}^{1}dt<n\frac{d~}{dt}F_{t}F_{t}%
^{n-1}I_{\xi}A_{t}+I_{\xi}\Delta AF_{t}^{n}>
\]
\[
+n(n+1)~d~\int_{0}^{1}dt<\Delta AF_{t}^{n-1}I_{\xi}A_{t}>
\]
but $I_{\xi}A_{t}=tI_{\xi}(\Delta A)+I_{\xi}\overline{A}$, then $I_{\xi
}(\Delta A)=\frac{d~}{dt}I_{\xi}A_{t}$ and therefore
\[
<n\frac{d~}{dt}F_{t}F_{t}^{n-1}I_{\xi}A_{t}+I_{\xi}\Delta AF_{t}^{n}%
>=\frac{d~}{dt}<F_{t}^{n}I_{\xi}A_{t}>
\]
This expression allows to integrate the first terms of the current,yielding
\begin{equation}
\ast j=<F^{n}I_{\xi}A>-<\bar{F}^{n}I_{\xi}\overline{A}>+n(n+1)~d~\int_{0}%
^{1}dt<\Delta AF_{t}^{n-1}I_{\xi}A_{t}>
\end{equation}
The first two terms of the second member vanish due to the E.O.M., then
\begin{equation}
\ast j=~d\mathbf{Q}_{\xi}%
\end{equation}
with
\begin{equation}
\mathbf{Q}_{\xi}=+n(n+1)\int_{0}^{1}dt<\Delta AF_{t}^{n-1}I_{\xi}A_{t}>
\end{equation}
The conserved charge is then
\begin{equation}
Q(\xi)=\int_{\partial\Sigma}\mathbf{Q}_{\xi}=+n(n+1)\int_{\partial\Sigma}%
\int_{0}^{1}dt<\Delta AF_{t}^{n-1}I_{\xi}A_{t}>
\end{equation}

>From this expression one gets the one for pure Chern-Simons by setting
$\overline{A}=0$, because the configuration $\overline{A}=0$ satisfies the E.O.M..

\subsection{Gauge Noether charges}

The variation of $A$ and $\overline{A}$ under gauge transformations is
\begin{equation}
\delta_{\lambda}A=-D\lambda~~,~~\delta_{\lambda}\overline{A}=-\overline
{D}\lambda,
\end{equation}
which implies
\begin{equation}
\delta_{\lambda}A_{t}=-D_{t}\lambda=-d\lambda-A_{t}\lambda+\lambda A_{t}%
\end{equation}
The E.O.M., which we assume are satisfied by both fields $A$ and $\overline
{A}$, are $<F^{n}T_{I}>=0$ and $<\bar{F}^{n}T_{I}>=0$. It follows that we can
read the form $\Theta$ appearing in the Noether theorem (see appendix) from
the expression for the variation
\begin{equation}
\Theta=n(n+1)\int_{0}^{1}dt<\Delta AF_{t}^{n-1}D_{t}\lambda>
\end{equation}
The form $\Omega$ is zero in this case, because the transgression is gauge
invariant. It follows that the conserved current is
\begin{equation}
\ast j_{\lambda}=-\Theta=-n(n+1)\int_{0}^{1}dt<\Delta AF_{t}^{n-1}D_{t}%
\lambda>
\end{equation}
Furthermore $\ast j_{\lambda}=d\mathbf{Q}_{\lambda}$ with
\begin{equation}
\mathbf{Q}_{\lambda}=n(n+1)\int_{0}^{1}dt<\Delta AF_{t}^{n-1}\lambda>
\end{equation}
because
\begin{equation}
d\mathbf{Q}_{\lambda}=n(n+1)\int_{0}^{1}dt<D_{t}[\Delta AF_{t}^{n-1}\lambda]>
\end{equation}
or
\begin{equation}
d\mathbf{Q}_{\lambda}=n(n+1)\int_{0}^{1}dt<\frac{d~}{dt}F_{t}F_{t}%
^{n-1}\lambda-\Delta AF_{t}^{n-1}D_{t}\lambda]>
\end{equation}
and, using $\frac{d~}{dt}F_{t}^{n-1}=\frac{1}{n}\frac{d~}{dt}F_{t}^{n}$ inside
the bracket $<~>$ we get
\begin{equation}
d\mathbf{Q}_{\lambda}=(n+1)<(F_{1}^{n}-F_{0}^{n})\lambda>-n(n+1)\int_{0}%
^{1}dt<\Delta AF_{t}^{n-1}D_{t}\lambda>
\end{equation}
where the first term of the second member is zero due to the E.O.M.

The conserved gauge charge is then
\begin{equation}
Q(\lambda)=\int_{\partial\Sigma}\mathbf{Q}_{\lambda}=n(n+1)\int_{\partial
\Sigma}\int_{0}^{1}dt<\Delta AF_{t}^{n-1}\lambda>
\end{equation}

This expression for $Q({\lambda})$ yields the one for a pure Chern-Simons
theory, by setting $\overline{A}=0$, because the configuration $\overline
{A}=0$ does satisfy the E.O.M., in agreement with the hypotheses of our derivation.

\subsection{\bigskip Algebra of the gauge charges}

If
\begin{equation}
\mathbf{Q}_{\lambda}=n(n+1)\int_{0}^{1}dt<\Delta AF_{t}^{n-1}\lambda>
\end{equation}
the algebra of the gauge charges is given by
\begin{equation}
\{\mathbf{Q}({\lambda}),\mathbf{Q}({\eta})\}:=\delta_{\eta}\mathbf{Q}%
({\lambda})
\end{equation}
To evaluate this expression we notice that under finite gauge transformations
\[
A\rightarrow g^{-1}[A+d]g~~,~~\overline{A}\rightarrow g^{-1}[\overline{A}+d]g
\]
hence $\Delta A\rightarrow g^{-1}\Delta A~g$, $A_{t}\rightarrow g^{-1}%
[A_{t}+d]g$, $F\rightarrow gFg^{-1}$ , $\bar{F}\rightarrow g\bar{F}g^{-1}$ and
$F_{t}\rightarrow gF_{t}g^{-1}$. Writing $g$ in the form $g=exp[-\lambda]$
with $\lambda=\lambda^{I}T_{I}$, in the case of an infinitesimal $\lambda$ we
recover the expressions for infinitesimal gauge transformations $\delta
_{\lambda}A=-D\lambda~~,~~\delta_{\lambda}\overline{A}=-\overline{D}\lambda$
and $\delta_{\lambda}A_{t}=-D_{t}\lambda=-d\lambda-A_{t}\lambda+\lambda A_{t}$.

To compute $\delta_{\eta}$ on $\mathbf{Q}({\lambda})$ it is easier to start
with a finite transformation $g=exp[-\eta]$ and then take the limit $\eta\ll
1$. We have
\[
<\Delta AF_{t}^{n-1}\lambda>\rightarrow<g^{-1}\Delta AF_{t}^{n-1}g\lambda>
\]
and, using the cyclic property of the trace
\[
<g^{-1}\Delta AF_{t}^{n-1}g\lambda>=<\Delta AF_{t}^{n-1}g\lambda g^{-1}>
\]
and using the infinitesimal form $g=1-\eta$ it results
\begin{equation}
\delta_{\eta}\mathbf{Q}({\lambda})=\mathbf{Q}({[\lambda,\eta]})
\end{equation}
or
\begin{equation}
\{\mathbf{Q}({\lambda}),\mathbf{Q}({\eta})\}=\mathbf{Q}({[\lambda,\eta]})
\end{equation}
Notice the absence of central charges in the second member, which was to be
expected, because central charges are characteristic of quasi-invariant
lagrangians and transgressions are truly invariant.

\section{Gravity and the transgression form}

In this Appendix we derive the explicit expression for the AdS group
transgression form given in the main text.

The AdS curvatures are
\begin{equation}
F=\frac{1}{2}\mathbf{R}J+TP~~~,~~~\overline{F}=\frac{1}{2}\overline
{\mathbf{R}}J+\overline{T}P
\end{equation}
where $R=d\omega+\omega^{2}$, $\overline{R}=d\overline{\omega}+\overline
{\omega}^{2}$ are the curvatures, $T^{a}=de^{a}+\omega_{~b}^{a}e^{b}=De^{a}$,
$\overline{T}^{a}=d\overline{e}^{a}+\overline{\omega}_{~b}^{a}\overline{e}%
^{b}=\overline{D}\overline{e}^{a}$ are the torsions, and $\mathbf{R}=R+e^{2}$,
$\overline{\mathbf{R}}=\overline{R}+\overline{e}^{2}$. Furthermore
\begin{equation}
F_{t}=\frac{1}{2}\mathbf{R}_{t}J+T_{t}P
\end{equation}
with
\begin{equation}
\mathbf{R}_{t}=t\mathbf{R}+(1-t)\overline{\mathbf{R}}-t(1-t)[\theta^{2}+E^{2}]
\end{equation}
where we define $\theta=\omega-\overline{\omega}=\Delta\omega$ (\emph{even
though $\theta$ is not the Second Fundamental Form for generic $\omega$ and
$\overline{\omega}$}) and $E=e-\overline{e}$. $\mathbf{R}_{t}$ can also be
written as
\begin{equation}
\mathbf{R}_{t}= tR+(1-t)\overline{R}-t(1-t)\theta^{2}+e_{t}^{2}%
\end{equation}
where $e_{t}=te+(1-t)\overline{e}$. Furthermore
\begin{equation}
T_{t}=tT+(1-t)\overline{T}-t(1-t)((\theta E))
\end{equation}
where the double parentheses stands for contractions, for instance
$((\omega^{2}))^{ab}\equiv\omega^{af}\omega_{f}^{~b}\equiv(\omega^{2})^{ab}$,
and $((\omega e))\equiv\omega^{af}e_{f}$.

The AdS transgression form must be the of the form
\begin{equation}
\mathcal{T}_{2n+1}^{AdS}=\kappa\int_{0}^{1}dt\epsilon(R+t^{2}e^{2}%
)^{n}e-\kappa\int_{0}^{1}dt\epsilon(\overline{R}+t^{2}\overline{e}^{2}%
)^{n}\overline{e}-dB_{2n}%
\end{equation}
where the boundary term $B_{2n}$ is what we intend to determine.

The variation of the bulk AdS CS form is
\begin{equation}
\delta\mathcal{C}_{2n+1}^{AdS}=\kappa\epsilon\mathbf{R}^{n}\delta
e+\kappa\epsilon n\mathbf{R}^{n-1}T\delta\omega+d\Xi
\end{equation}
where the boundary contribution is
\begin{equation}
\Xi=-\kappa n\int_{0}^{1}dt\epsilon(R+t^{2}e^{2})^{n-1}e\delta\omega
\end{equation}

The variation of the transgression must then be
\begin{equation}
\delta\mathcal{T}_{2n+1}^{AdS}=\kappa\epsilon\mathbf{R}^{n}\delta
e+\kappa\epsilon n\mathbf{R}^{n-1}T\delta\omega-\kappa\epsilon\overline
{\mathbf{R}}^{n}\delta\overline{e}-\kappa\epsilon n\overline{\mathbf{R}}^{n-1}
\overline{T}\delta\overline{\omega}+d\Xi-d\overline{\Xi}+d[\delta B_{2n}]
\end{equation}
But
\begin{eqnarray}
\kappa\epsilon\mathbf{R}^{n}\delta e+\kappa\epsilon n\mathbf{R} ^{n-1}%
T\delta\omega-\kappa\epsilon\overline{\mathbf{R}}^{n}\delta\overline{e}%
-\kappa\epsilon n\overline{\mathbf{R}}^{n-1}\overline{T}\delta\overline
{\omega}=\nonumber\\
=(n+1)<F^{n}\delta A>-(n+1)<\overline{F}^{n}\delta\overline{A}>
\end{eqnarray}
So, except for an irrelevant closed form, it must hold that
\begin{equation}
\Theta_{2n}=\Xi-\overline{\Xi}+\delta B_{2n}%
\end{equation}
Our goal is to find $B_{2n}$. To that end we notice that $\Xi$ contains only
$\delta\omega$ while $\overline{\Xi}$ contains only $\delta\overline{\omega}$,
therefore the coefficients of $\delta e$ and $\delta\overline{e}$ on
$\Theta_{2n}$ and $B_{2n}$ must be the same. We will exploit this fact to
integrate the variations.

For AdS
\begin{equation}
\Theta_{2n}=-\kappa n\int_{0}^{1}dt\epsilon\{[\theta\mathbf{R}_{t}%
^{n-1}]\delta e_{t}+[E\mathbf{R}_{t}^{n-1}+(n-1)\theta\mathbf{R}_{t}%
^{n-2}T_{t}]\delta\omega_{t}\}
\end{equation}
But
\begin{equation}
\epsilon\mathbf{R}_{t}^{n-1}\delta e_{t}=\epsilon\lbrack\xi_{t}(\omega
,\overline{\omega})+e_{t}^{2}]^{n-1}\delta e_{t}%
\end{equation}
with
\begin{equation}
\xi_{t}(\omega,\overline{\omega})=tR+(1-t)\overline{R}-t(1-t)\theta^{2}%
\end{equation}
Expanding we get
\begin{equation}
\epsilon\mathbf{R}_{t}^{n-1}\delta e_{t}=\epsilon\left\{  \sum_{k=0}%
^{n-1}C_{k}^{n-1}\xi_{t}^{n-1-k}e_{t}^{2k}\delta e_{t}\right\}  =\epsilon
\delta¨_{(e_{t})}\left\{  \sum_{k=0}^{n-1}\frac{C_{k}^{n-1}}{2k+1}\xi
_{t}^{n-1-k}e_{t}^{2k+1}\right\}
\end{equation}
where we used the symbol $\delta_{(e_{t})}$ in the last member to indicate
that only the vielbeins are varied there. We can then write
\begin{eqnarray}
\epsilon\mathbf{R}_{t}^{n-1}\delta e_{t}  & =&\delta_{(e_{t})}\left\{
\int_{0}^{1}ds\epsilon\left[  \sum_{k=0}^{n-1}C_{k}^{n-1}\xi_{t}^{n-1-k}%
s^{2k}e_{t}^{2k}\right]  e_{t}\right\} \nonumber\\
& =&\epsilon\delta_{(e_{t})}\left\{  \int_{0}^{1}ds~\epsilon\theta e_{t}%
[\xi_{t}+s^{2}e_{t}^{2}]^{n-1}\right\} \nonumber
\end{eqnarray}
where again only the vielbeins are varied. It follows that
\begin{equation}
B_{2n}=\kappa n\int_{0}^{1}dt\int_{0}^{1}ds~\epsilon~\theta e_{t}\left\{
tR+(1-t)\overline{R}-t(1-t)\theta^{2}+s^{2}e_{t}^{2}\right\}  ^{n-1}%
\end{equation}

It may worry the reader that a contribution to $B_{2}n$ depending only on
$\omega$ and $\overline{\omega}$ could have been missed in this approach,
however looking at $\Theta_{2n}$ above it is clear that such a contribution
does not exist, as every term has a dependence on $e$ or $\overline{e}$.

\section{Conserved charges for black holes with a reference background}

In order to evaluate the Noether charge associated to the time like Killing
vector $\xi=\frac{\partial~}{\partial t}$ for two black hole configurations
with parameters $\sigma$ and $\overline{\sigma}$ respectively the relevant non
vanishing ingredients \cite{scan,motz} are\footnote{in this Appendix, as in
the previous one $\theta=\omega-\bar{\omega}$, even though for generic
connections $\theta$ is not he Second Fundamental Form. We also use the
$\mathbf{R}_{t}$ notation of the previous Appendix.}
\begin{eqnarray}
\theta^{1m}  & =&-(\Delta-\overline{\Delta})\tilde{e}^{m}~, ~~~~~~~~~~~~~~~~~~(\theta^{2}%
)^{mn}=-(\Delta-\overline{\Delta})\tilde{e}^{m}\tilde{e}^{n}\nonumber\\
I_{\xi}A_{t}  & =&[t\Delta+(1-t)\overline{\Delta}]P_{0}+rJ_{01}~,~~(e_{t}%
^{2})^{mn}=r^{2}\tilde{e}^{m}\tilde{e}^{n}\nonumber\\
R^{mn}  & =&(1-\Delta^{2})\tilde{e}^{m}\tilde{e}^{n}~,~~~~~~~~~~~~~~~~~~~\bar{R}^{mn}%
=(1-\overline{\Delta}^{2})\tilde{e}^{m}\tilde{e}^{n}\nonumber\\
R^{0m}  & =&-\Delta
r~dt\tilde{e}^{m}~,~~~~~~~~~~~~~~~~~~~~~~~~\bar{R}^{0m}=-\overline{\Delta
}r~dt\tilde{e}^{m}%
\end{eqnarray}
Also $\Delta=(1-\sigma+r^{2})^{\frac{1}{2}}$ with $\sigma=(2Gm+1)^{\frac{1}%
{n}}$ and $\overline{\Delta}=(1-\overline{\sigma}+r^{2})^{\frac{1}{2}}$ with
$\overline{\sigma}=(2G\overline{m}+1)^{\frac{1}{n}}$. In the previous
expressions $m$ and $\overline{m}$ are just certain constants of integration of
the solutions, which turn to be closely related to the energy coming from the
thermodynamics and the one coming from the Noether charge, justifying in
retrospect to call those parameters the 'masses' of the black holes.

We will need the components of $\mathbf{R}_{t}=tR+(1-t)\bar{R}-t(1-t)\theta
^{2}+e_{t}^{2}$, where $e_{t}=te+(1-t)\bar{e})$, with group indices $mn$.
Those are
\begin{equation}
\left(  tR+(1-t)\bar{R}-t(1-t)\theta^{2}+e_{t}^{2}\right)  ^{mn}=\left\{
1-[t\Delta+(1-t)\overline{\Delta}]^{2}+r^{2}\right\}  \tilde{e}^{m}\tilde
{e}^{n}%
\end{equation}
The charge coming from eq.(\ref{Qdiff})is in this case
\begin{equation}
Q(\frac{\partial~}{\partial t})=\kappa n\int_{\partial\Sigma}\int_{0}%
^{1}dt~2\epsilon_{01m_{1}m_{2}...m_{2n-1}}[I_{\xi}A_{t}]^{0}\theta^{1m_{1}%
}\mathbf{R}_{t}^{m_{2}m_{3}}...\mathbf{R}_{t}^{m_{2n-2}m_{2n-1}}%
\end{equation}
where we used that the only non vanishing components of $\Delta A$ with
support in the spatial boundary are $\theta^{1m}$, so the index 1 is
necessarily there, while the index 0 must then be in $I_{\xi}A_{t}$, which
must therefore contain the only generator $P_{a}$. Inserting the expressions
for the terms of this equation and integrating in the spatial boundary
$\partial\Sigma\equiv S^{d-2}$ we get
\begin{eqnarray}
Q(\frac{\partial~}{\partial t})  & =& -\kappa2n\Omega_{d-2}\int_{0}%
^{1}dt\{(\Delta-\overline{\Delta})[t\Delta+(1-t)\overline{\Delta}%
]\times\nonumber\\
& & \times\lbrack1-(t\Delta+(1-t)\overline{\Delta})^{2}+r^{2}]^{n-1}\}
\end{eqnarray}
where $\Omega_{d-2}$ is the volume of the sphere of dimension $d-2$ resulting
from the integration of the angular variables. The integral in the parameter
$t$ can be done trough the substitution $u=1-(t\Delta+(1-t)\overline{\Delta
})^{2}+r^{2}$ and the result is
\begin{equation}
Q(\frac{\partial~}{\partial t})=\kappa(d-2)!\Omega_{d-2}[u^{n}]\mid
_{1-\overline{\Delta}^{2}+r^{2}}^{1-\Delta^{2}+r^{2}}%
\end{equation}
Notice that $1-\Delta^{2}+r^{2}=\sigma$ and $1-\overline{\Delta}^{2}%
+r^{2}=\overline{\sigma}$. The result is
\begin{equation}
Q(\frac{\partial~}{\partial t})=\kappa(d-2)!\Omega_{d-2}[\sigma^{n}%
-\overline{\sigma}^{n}]
\end{equation}
Using the expressions for $\sigma$ and $\overline{\sigma}$, that $\kappa
=\frac{1}{2G(d-2)!\Omega_{d-2}}$ we get
\begin{equation}
Q(\frac{\partial~}{\partial t})=m-\overline{m}=E-\bar{E}%
\end{equation}
Particular cases of this expression are the zero mass black hole (with
$\overline{m}=0$) and AdS (with $\overline{m}=-\frac{1}{2G}$). The AdS mass
can be thought as a vacuum or Casimir energy.\newline

A similar calculation yields the charge for topological black holes.
Topological black holes are labeled by the parameter $\gamma$ which can take
the values 1, 0 or -1. The line element is
\begin{equation}
ds^{2}=-\Delta^{2}dt^{2}+\frac{1}{\Delta^{2}}dr^{2}+r^{2}d\Sigma_{d-2}^{2}%
\end{equation}
where $d\Sigma_{d-2}^{2}$ is the line element of the $d-2$ sphere ($\gamma=1$,
the case just considered), plane ($\gamma=0$) or hyperboloid ($\gamma=-1$)
and
\begin{eqnarray}
\Delta^{2}  & =& \gamma-\sigma+r^{2}\nonumber\\
\sigma & =& \left(  2G\mu+\delta_{1,\gamma}\right)  ^{\frac{1}{n}}%
\end{eqnarray}
where $\Sigma_{d-2}$ stands for the volume of the corresponding $d-2$
dimensional manifold and $\mu$ is a parameter or integration constant which,
after the thermodynamics and Noether charge computation of the energy, will be
regarded as a energy or mass density.

The calculation of Noether charge is analogous. We now have
\begin{equation}
\left(  \mathbf{R} _{t}\right)  ^{mn}= \left\{  \gamma-[t\Delta+(1-t)\overline
{\Delta}]^{2}+r^{2}\right\}  \tilde{e}^{m}\tilde{e}^{n}%
\end{equation}
The charge is
\begin{equation}
Q(\frac{\partial~}{\partial t}) =\kappa(d-2)!\Sigma_{d-2} [u^{n}%
]\mid_{\overline{\sigma}}^{\sigma}=\kappa(d-2)!\Sigma_{d-2} [\sigma
^{n}-{\overline{\sigma}}^{n}]
\end{equation}
where now $\partial\Sigma\equiv\Sigma^{d-2}$. Using the expressions for
$\sigma$ and $\overline{\sigma}$ and $\kappa=\frac{1}{2G(d-2)!\Omega_{d-2}}$
we get
\begin{equation}
Q(\frac{\partial~}{\partial t}) =\frac{\Sigma_{d-2}}{\Omega_{d-2}}%
[\mu-\overline{\mu}]=E-\bar{E}%
\end{equation}
which implies that the parameter $\mu$ is a sort of energy density.

We will furthermore consider the $\overline{A}$ configuration for with the
horizon radius is zero $\overline{r}_{+}=0$, so that $\overline{\sigma}%
=\gamma$. For $\gamma=0,-1$ (the $\gamma=1$ case was studied above) we get
\begin{equation}
Q(\frac{\partial~}{\partial t})=\frac{\Sigma_{d-2}}{\Omega_{d-2}}(\mu
-\frac{\gamma^{n}}{2G})
\end{equation}
Here $\frac{\gamma^{n}}{2G}$ can be interpreted as a vacuum or Casimir energy density.


\begin{thebibliography}{99}                                                                                               %

\bibitem {QMFS}R. Troncoso and J. Zanelli, \textit{Chern-Simons
supergravities with off-shell local superalgebras}, in \textit{Black holes and
the structure of the universe, }119-145. Santiago, Chile, 18-20 Aug 1997. C.
Teitelboim and J. Zanelli (Eds.), World Scientific, Singapore, 2000.
hep-th/9902003.

\bibitem {zanelli2} J. Zanelli, \textit{Lecture notes on Chern-Simons (super-) gravities},
hep-th/0502193.

\bibitem {aschwarz}A. Schwarz, Lett. Math. Phys. \textbf{2}(1978)247.

\bibitem {deser}S. Deser, R. Jackiw and S. Templeton, Phys. Rev. Lett.
\textbf{48}(1983)975; Ann. Phys. NY \textbf{140}(1984)372.

\bibitem {witten-cs}E. Witten, Comm. Math. Phys. \textbf{121}(1989)351.

\bibitem {vannieuwen}P. Van Nieuwenhuizen, Phys.~Rev.~ \textbf{D32}(1985) 872.

\bibitem {achu}A. Ach\'ucarro and P.K. Townsend, Phys.~Lett.~\textbf{B180}(1986)89.

\bibitem {witten1}E. Witten, Nucl.~Phys.~\textbf{311B}(1988)46;
Nucl.~Phys.~\textbf{323B}(1989)113.

\bibitem {chamseddine}A.H. Chamseddine, Phys.~Lett.~\textbf{B233}(1989)291.

\bibitem {troncoso3}R. Troncoso and J. Zanelli, Class. Quan. Grav.
\textbf{17}(2000)4451, hep-th/9907109.

\bibitem {chamseddine-Nuc}A.H. Chamseddine, Nucl.~Phys.~\textbf{346B}(1990)213.

\bibitem {troncoso1}R. Troncoso and J. Zanelli, Phys.~Rev.~ \textbf{D58}%
(1998)101703, hep-th/9710180.

\bibitem {troncoso2}R. Troncoso and J. Zanelli, Int. Jour. Theor. Phys.
\textbf{38}(1999)1181, hep-th/9807029.

\bibitem {banados1}M. Ba\~{n}ados, R. Troncoso and J. Zanelli, Phys.~Rev.~
\textbf{D54}(1996)2605.

\bibitem {HOT}M.~Hassaine, R.~Olea and R.~Troncoso, Phys.\ Lett.\ B
\textbf{599}, 111 (2004), hep-th/0210116.

\bibitem {hassaine}M. Hassaine, R. Troncoso and J. Zanelli, Phys.Lett.
\textbf{B596}(2004)132, hep-th/0306258; PoS \textbf{WC2004}(2005)006, hep-th/0503220.

\bibitem {townsend1}P.K. Townsend, Phys.~Lett.~\textbf{B350} 1995)184, hep-th/9501068.

\bibitem {hull1}C.~Hull and P.K.~Townsend, Nucl.~Phys.~\textbf{348B}(1995)109.

\bibitem {witten2}E. Witten, Nucl.~Phys.~\textbf{443B}(1995)85.

\bibitem {townsend2}P.K. Townsend, \textit{Four Lectures on M-Theory}, in
\textit{Proceedings of ICTP Summer School on High Energy Physics and
Cosmology}, Trieste (June 1996), hep-th/9612121.

\bibitem {horava}P. Horava, Phys. Rev. \textbf{D59}(1999)046004, hep-th/9712130.

\bibitem {banados2}M.~Banados, Phys.\ Rev.\ Lett.\ \textbf{88}, 031301 (2002),
hep-th/0107214.

\bibitem {nastase}H.~Nastase, \textit{Towards a Chern-Simons M theory of}
$OSp(1|32)xOSp(1|32)$, hep-th/0306269.


\bibitem {stora}R. Stora, \textit{Algebraic Structure of Chiral Anomalies}, in
\textit{Recent Progress in Gauge Theories}, H. Lehmann ed., NATO ASI Series,
(Plenum, NY, 1984).

\bibitem {zumino}B. Zumino, \textit{Chiral Anomalies and Differential
Geometry}, in \textit{Relativity, Groups and Topology II}, B.S. De Witt and R.
Stora eds., (North Holland, Amsterdam, 1984).

\bibitem {manes}J. Ma\~{n}es, R. Stora and B. Zumino, Comm. Math. Phys.
\textbf{102}(1985)157.

\bibitem {alvarez}L. Alvarez-Gaum\'{e} and P. Ginsparg,
Ann.~of~Phys.~\textbf{161}(1985)423.

\bibitem {chern}S. S. Chern, \textit{Complex Manifolds without Potential
Theory}, 2nd Ed., (Springer, Berlin, 1979).

\bibitem {nakahara}M. Nakahara, \textit{Geometry, Topology and Physics}, (IOP,
Bristol, 1991).

\bibitem {potsdam}R. Aros, M. Contreras, R. Olea, R. Troncoso and J.
Zanelli,\textit{Charges in 2+1 Dimensional Gravity and Supergravity}, presented
at the \textit{Strings'99 Conference}, Potsdam, Germany, July 1999.

\bibitem {motz}P. Mora, R. Olea, R. Troncoso and J. Zanelli, JHEP\textbf{06}%
(2004)036, hep-th/0405267.

\bibitem {mn}P. Mora and H. Nishino, Phys. Lett. \textbf{B482}(2000)222, hep-th/0002077.

\bibitem {mora}P. Mora, Nucl. Phys. \textbf{594B}(2001)229, hep-th/0008180.

\bibitem {BFF2003}A. Borowiec, M.~Ferraris and M.~Francaviglia, J. Phys. A
\textbf{36}, 2589 (2003), hep-th/0301146.

\bibitem {BFFF2005}A. Borowiec, L. Fatibene, M. Ferraris and M. Francaviglia,
\emph{Covariant Lagrangian Formulation of Chern-Simons and BF Theories}, hep-th/0511060.

\bibitem {IRS}F. Izaurieta, E. Rodriguez and P. Salgado, \emph{On
Transgression Forms and Chern--Simons (Super)gravity}, hep-th/0512014.

\bibitem {tesis}P. Mora, \textit{Formas de Transgresi\'on como Principio
Unificador en Teor\'{\i}a de Campos}, Ph. D. Thesis, Universidad de la
Rep\'{u}blica, Uruguay, (2003), hep-th/0512255.

\bibitem {BTZentropy}M. Ba\~{n}ados, C. Teitelboim and J. Zanelli, Phys. Rev.
Lett., \textbf{72} (1994) 957..

\bibitem {scan}J. Cris\'{o}stomo, R. Troncoso and J. Zanelli, Phys. Rev.
\textbf{D62}(2000)084013.

\bibitem {topo2}R. Aros, R. Troncoso and J. Zanelli, Phys. Rev. \textbf{D63}(2001)084015.

\bibitem {Silva}S. Silva, Nucl. Phys. \textbf{B558} (1999) 391, hep-th/9809109.

\bibitem {sarda}G. Sardanashvily, \textit{Gauge conservation laws in
higher-dimensional Chern-Simons models}, hep-th/0303059;
\textit{Energy-momentum conservation in higher-dimensional Chern-Simons
models}, hep-th/0303148.

\bibitem {Thesis+Preprint+BFF+IRS}The expressions for the conserved charges
(\ref{Qdiff}) and (\ref{Qgauge}) for a transgression form have been written in
Pablo Mora, Ph. D. Thesis, Universidad de la Rep\'{u}blica (2003)
\cite{tesis}, as well as in the preprint P. Mora, R. Olea, R. Troncoso and J.
Zanelli CECS-PHY-04/13. This has also been recently discussed in Refs.
\cite{BFFF2005} and in \cite{IRS}.

\bibitem {Lovelock}D. Lovelock, J.Math.Phys. \textbf{12} (1971) 498.

\bibitem {dimensionally}M. Ba\~{n}ados, C. Teitelboim and J. Zanelli, Phys.
Rev. Lett. 69(1992)1849; Phys. Rev. \textbf{D49}(1994)975.

\bibitem {topo1}R.G. Cai and K.S. Soh, Phys. Rev. \textbf{D59} (1999)044013

\bibitem {gh}G.W. Gibbons and S.W. Hawking, Phys. Rev. \textbf{D15}(1977)2753.

\bibitem {hp}S.W. Hawking and D.N. Page, Commun. Mat. Phys. \textbf{87}(1983)577.

\bibitem {aroscargas}R. Aros, M. Contreras, R. Olea, R. Troncoso and J.
Zanelli, Phys. Rev. Lett. 84(2000)1647; Phys. Rev. \textbf{D62} (2000)044002.

\bibitem {yang-mills}\textit{``It may appear that $B_{\mu}$ can be introduced
as an auxiliary quantity to accomplish invariance, but need not be regarded as
a field variable by itself. It is to be emphasized that such a procedure
violates the principle of invariance. Every quantity that is not a pure numeral
(like 2, or $M$, or any definite representation of the $\gamma$ matrices)
should be regarded as a dynamical variable, and should be varied in the
Lagrangian to yield an equation of motion. Thus the quantities $B_{\mu}$ must
be regarded as independent fields"}.C.N. Yang and R.L. Mills, Phys. Rev.
\textbf{96}(1954)191.

\bibitem {Mardones-z}A. Mardones and J. Zanelli, Class. Quant. Grav.\textbf{8}
(1991) 1545.

\bibitem {maldacena2}J. Maldacena, Adv. Theor. Math. Phys. 2(1998)231; Int. J.
Theor. Phys. 38(1999)1113, hep-th/0309246.

\bibitem{bertlmann}R. Bertlmann, \textit{Anomalies in Quantum Field Theory},
(Oxford U.P., Oxford, 1996).

\end{thebibliography}
\end{document}